\documentclass[article, shortnames,nojss]{jss}

\usepackage{thumbpdf,lmodern}
\usepackage{amsmath}
\usepackage{soul}
\usepackage{algorithm}
\usepackage{algorithmic}
\usepackage{multirow}
\usepackage{subcaption}
\usepackage{natbib}

\DeclareMathOperator*{\argmax}{argmax} 
\usepackage{framed}
\usepackage{subfiles}
\newcommand{\class}[1]{`\code{#1}'}
\newcommand{\fct}[1]{\code{#1()}}

\newcommand{\vX}{\mbox{\boldmath $X$}}
\newcommand{\vbeta}{\mbox{\boldmath $\beta$}}
\newcommand{\vx}{\mbox{\boldmath $x$}}
\newcommand{\vgamma}{\mbox{\boldmath $\gamma$}}
\newcommand{\vy}{\mbox{\boldmath $y$}}

\author{Qianxiang Zang\\University of Ottawa
    \And Chen Xu\\University of Ottawa
   \And Kelly Burkett\\University of Ottawa}
\Plainauthor{Qianxiang Zang, Chen Xu, Kelly Burkett}

\title{\pkg{SMLE}: An \proglang{R} Package for Joint Feature Screening in Ultrahigh-dimensional GLMs}

\Plaintitle{SMLE: An R Package for Joint Feature Screening in Ultrahigh-dimensional GLMs}

\Shorttitle{\proglang{R} package \pkg{SMLE}}

\Abstract{
  The sparsity-restricted maximum likelihood estimator (SMLE) has received considerable attention for feature screening in ultrahigh-dimensional regression. SMLE is a computationally convenient method that naturally incorporates the joint effects among features in the screening process.
  We develop a publicly available \proglang{R} package \pkg{SMLE}, which provides a user-friendly environment to carry out SMLE in generalized linear models. In particular, the package includes functions to conduct SMLE-screening and post-screening selection using SMLE with popular selection criteria such as AIC and (extended) BIC. The package gives users the flexibility in controlling a series of screening parameters and accommodates both numerical and categorical feature input. The usage of the package is illustrated on extensive numerical examples, where the promising performance of SMLE is well observed. 
}

\Plainkeywords{Ultrahigh-dimensional data, Joint feature screening, Iterative hard-thresholding, Generalized linear models, EBIC}

\Address{
  Qianxiang Zang, Chen Xu, Kelly Burkett\\
  Department of Mathematics and Statistics\\
  Faculty of Science\\
  University of Ottawa\\
  E-mail:  \email{qzang023@uottawa.ca}, \email{cx3@uottawa.ca}, \email{kburkett@uottawa.ca}\\

}

\begin{document}
\section[Introduction]{Introduction} \label{sec:intro}

In modern scientific research, it is common to encounter ultrahigh-dimensional datasets with a huge number of features.  
For example, geneticists often need to measure thousands to hundreds of thousands of genes in the hope of discovering those that influence an observable trait; an Internet firewall may scan millions of keywords on data packets in order to determine their security risk. While ultrahigh-dimensional data bring rich resources to explore many unknown areas, they pose simultaneous challenges of computational cost, statistical accuracy, and algorithmic stability for classic statistical methods \citep{Fan2009}.

When the number of features is huge, it is often reasonable to assume that only a handful of them are relevant to the analysis. In a regression setting, this amounts to assuming that most predictors in an ultrahigh-dimensional model have no effect on the response (i.e. the regression coefficient is zero). With this sparsity assumption, one natural strategy is to screen most irrelevant features out before a more elaborate analysis is conducted. This pre-processing procedure is referred to as feature screening. With dimensionality reduced from high to low, analytical difficulties can be reduced drastically.

In recent years, much research has been done on feature screening. \citet{Fan+Lv:2008} proposed to screen features based on their marginal Pearson correlations with the response; they referred to this procedure as sure independence screening (SIS) and justified its theoretical effectiveness for linear models. \citet{FanSong} extended SIS to generalized linear models (GLMs). In the same spirit, \citet{Zhu2011} proposed a sure independent ranking and screening (SIRS) based on the conditional distribution of the response given each feature. \citet{Li2012} developed a model-free sure independence screening based on the distance correlation (DC-SIS). \citet{WuandYin2015} proposed a distribution function screening by testing the independence between the response and each feature. \citet{Zhou2019ModelFree} proposed a robust screening method for features containing extreme values. \citet{li2020distributed} proposed a distributed screening framework for the divide-and-conquer setup. The list above is certainly far from complete; readers may refer to \citet{Liu2015overview} for a selective overview on feature screening.

In the literature, most screening methods are developed based on the marginal effects of features on the response. 
Despite the convenience of implementation, these methods are often found to be unreliable in practice, as the joint effects among features are ignored. Features with significant joint effects but showing weak marginal effects are likely to be wrongly screened out. To tackle this issue, \citet{Fan+Lv:2008} suggested applying SIS iteratively (ISIS) with a smaller number of features retained in each round. \citet{Wang2009Forward} suggested using classic forward regression for screening purposes. These strategies help to incorporate some feature joint effects in the screening process. However, they are usually at a high computational cost, which can be unfavorable in many applications. 

Starting from a different angle,
\citet{Chen+Chen:2014} proposed a joint feature screening method via the sparsity restricted maximum likelihood estimator (SMLE). With a $L_0$ penalty specifying the number of features allowed in the model, the method attempts to roughly estimate a handful of the most significant coefficients from the full model while setting all other coefficients to zero. Since the estimation is carried out on the full model, the resulting sparse estimator readily serves as a feature screener, which naturally takes the joint effects among features into account. The SMLE method can be efficiently implemented by an iterative hard-thresholding algorithm (IHT), which does not involve any complicated numerical operations such as matrix inversion. Each IHT iteration increases the value of the sparsity-constrained joint likelihood and thereby provides an improved sparse solution, which eventually leads to a reliable screening result. \citet{Chen+Chen:2014} further justified the sure screening property of SMLE in ultrahigh-dimensional GLMs and proved the convergence of the IHT algorithm. Since SMLE, the idea of using a $L_0$ constraint in feature screening has attracted considerable attention. \citet{Yang2016Feature} and \cite{Yang2018Feature} extended SMLE to the ultrahigh-dimensional Cox model.
\citet{Qu2020} developed a $L_0$-based screening method for ultrahigh-dimensional quantile regression.

In this paper, we provide a publicly available \proglang{R} package \pkg{SMLE}\footnote{Available from the Comprehensive \proglang{R}
Archive Network (CRAN) at \url{https://cran.r-project.org/web/packages/SMLE}}, which gives a user-friendly environment to carry out SMLE in ultrahigh-dimensional GLMs including linear, logistic, and Poisson models.  The package makes use of the \fct{crossprod} function to handle ultrahigh-dimensional matrix products and includes a well-tuned main function to efficiently conduct SMLE-screening based on the IHT algorithm.
%
%
With the package, we are able to repeat the same numerical experiments in \citet{Chen+Chen:2014} at a significantly reduced time cost in comparison with the original code provided by the authors. 
In the package, we extend SMLE by permitting both numerical and categorical features in the screening, where the categorical features can be automatically identified and encoded by a user-selected method. Moreover, combined with popular selection criteria such as AIC or (extended) BIC, we propose a SMLE-based selection method, which helps to further identify the relevant features after screening. This post-screening selection method can be conveniently conducted within the main screening function or can be used independently on a user-supplied dataset.
We illustrate the usage of the package via extensive numerical examples. The promising performance of the package is observed in comparison with \pkg{SIS} \citep{Saldana+Feng:2018,RSIS}, which is the \proglang{R} package implementing both SIS and ISIS.

The rest of the paper is organized as follows. In Section 2, we give a brief overview of SMLE and the IHT algorithm. 
In Section 3, we discuss implementation details of the \pkg{SMLE} package. 
In Section 4, we illustrate the usage of the package using extensive numerical examples and compare its performance with the \proglang{R} package \pkg{SIS}. We conclude the paper in Section 5 with a few remarks.

\section{The SMLE method} \label{sec:models}

\subsection{Notation and problem setup}

Suppose the data $\{(y_{i}, \boldsymbol{x}_{i}), i=1,\ldots,n \}$ are collected independently from $(Y, \boldsymbol{x})$, where $Y$ is a response variable and $\boldsymbol{x}=(x_{1}, \ldots, x_{p})$ is a $p$-dimensional covariate (feature) vector.
We postulate a GLM between $Y$ and $\boldsymbol{x}$ as follows. Conditioning on $\boldsymbol{x}$, the distribution of $Y$ is assumed to belong to an exponential family taking the form
\begin{equation*}  \label{expf}
f(y; \theta)=\exp(\theta y - b(\theta) + c(y) ),
\end{equation*}
where $\theta$ is the natural or canonical parameter and $b(.)$, $c(.)$ are two known functions. Under the canonical link, $\boldsymbol{x}$ influences $Y$ in the form of a linear combination 
$$\theta = \boldsymbol{x}\boldsymbol{\vbeta},$$
where $\boldsymbol{\vbeta}=(\beta_{1}, \ldots, \beta_{p})^{T}$ is a $p$-dimensional regression coefficient. Popular GLMs with canonical links include the normal linear model, the logistic model, and the log-linear Poisson model \citep{GLMs}.

Under the GLM framework, the effect of each feature $x_j$ on the response $Y$ is characterized by the size of the corresponding regression coefficient $\beta_j$. In applications,  when the number of features $p$ is large, it is often believed that only a small number of the features in $\boldsymbol{x}$ contribute to the variation in $Y$, which leads to an idealistic assumption that $\vbeta$ contains many zero elements. With this sparsity assumption, only features with non-zero coefficients are considered to be relevant in explaining the variation of $Y$. The goal of feature screening is to identify and remove most of the irrelevant features, so that a more elaborate analysis can be conducted only on the features most likely to be related to $Y$.

\subsection{The SMLE-screening and IHT algorithm} \label{subsec:smle}

The idea of SMLE is simple. When $p$ is overly large and most model coefficients are assumed to be zero, it is probably not wise to estimate the entire $\boldsymbol{\vbeta}$ from scratch. Instead, it is reasonable to consider just estimating some of the coefficients while setting the others to zero from the beginning. This leads to a sparsity-restricted estimation, which readily serves for feature screening.

Specifically, under the GLM given in Section 2.1, the log-likelihood function of $\boldsymbol{\vbeta}$ is given by
\begin{equation*}  \label{log-lh}
l(\boldsymbol{\vbeta}) = \sum_{i=1}^{n} [y_{i} \cdot \boldsymbol{x}_{i} \boldsymbol{\vbeta} - b( \boldsymbol{x}_{i} \boldsymbol{\vbeta}) ].
\end{equation*}
With a user-specified sparsity $k < p$, the SMLE estimator is defined by
\begin{equation} \label{eq:SMLE}
\hat{\vbeta}_{k}=\argmax\limits_{\vbeta}  l(\vbeta)\quad  \text{subject to}\quad ||\vbeta||_0 \leq k,
\end{equation}  
where $\|.\|_0$ is the vector $L_0$ norm indicating the number of non-zero elements in that vector. Clearly, $\hat{\vbeta}_{k}$ is designed to set all but the most significant $k$ coefficients to be zero; this amounts to identifying $k$ important features supported most by the joint likelihood. When $p$ is large and $k$ is chosen to be much smaller than $p$, $\hat{\vbeta}_{k}$ can be viewed as a feature screener, which naturally takes the joint effects among features into account.  The idea of SMLE has similarities with the use of $L_{0}$-regularized techniques in image processing,
where sparsity-constrained least-squares methods are frequently used to construct sparse representations for high-resolution images \citep{Donoho2006,Blumeth2009}.

While SMLE is conceptually simple, carrying out problem  
(\ref{eq:SMLE}) can be numerically challenging, as it is a high-dimensional combinatorial optimization. However, since our goal is feature screening,  finding the global solution to (\ref{eq:SMLE}) is not a major concern. In fact, it would suffice if we can obtain a good local solution, which helps to retain all relevant features.

In this spirit, \citet{Chen+Chen:2014} proposed an iterative hard-thresholding algorithm (IHT) to approximately solve (\ref{eq:SMLE}). The idea is as follows. With a $\vgamma$ close to $\vbeta$, one can approximate $l(\boldsymbol{\vbeta})$ by a surrogate function
\begin{equation} \label{eq:h-function}
h(\vbeta, \vgamma) =  l(\vgamma)+( \vbeta - \vgamma)^{T} l'(\vgamma) - (u/2)|| \vbeta - \vgamma ||_2^2,
\end{equation}
where $l'(\vgamma)= \partial l(\vgamma) / {\partial \vgamma}$, $\|.\|_2$ indicates the $L_2$ (Euclidean) norm, and $u>0$ is a scale parameter.  The first two terms in (\ref{eq:h-function}) match the Taylor's expansion of $l(\vbeta)$ at $\vbeta = \vgamma$, and the third term is introduced as a regularization term to enhance the convexity. 

The reason for introducing $h(\vbeta, \vgamma)$ is that it is separable in the components of $\vbeta$ and thus serves as a surrogate of $l(\vbeta)$ to conveniently carry out the sparsity-restricted maximization over $\vbeta$. Specifically, with an initial value $\vbeta^{(0)}$, we can seek a local solution of (\ref{eq:SMLE}) via the following iterative procedure.

\begin{equation} \label{eq:IHT}
\vbeta^{(t+1)}= \argmax\limits_{\vbeta} h(\vbeta,\vbeta^{(t)})\quad \text{subject}\  \text{to}\quad ||\vbeta||_0<k.
\end{equation}

Let $\vy=(y_{1}, \ldots, y_{n})^{T}$ and $\vX=(\vx_{i}^{T}, \ldots, \vx_{n}^{T})^{T}$. Because of the additivity of $\vbeta$ in $h$, the optimization in (\ref{eq:IHT}) takes a unified  form
\begin{equation}\label{eq:IHT-2}
 \min_{\vbeta} \frac{1}{2} \left\|\vbeta - u^{-1}[u \vbeta^{(t)}  + \vX^{T}\vy -  \vX^{T}b'(\vX\vbeta^{(t)})] \right\|_{2}^{2}   \quad \mbox{subject to} \ \|\vbeta\|_{0} \leq k,
\end{equation}
where $b'$ is the derivative of the $b(.)$ function depending on the choice of GLM \footnote{For normal linear model, $b'(\theta)= \theta $; for logistic model, $b'(\theta)=\exp{(\theta)}/(1 + \exp{(\theta}))$; for Poisson model, $b'(\theta)=\exp{(\theta)}.$}. Obviously, if (\ref{eq:IHT-2}) does not come with the sparsity constraint, its solution should be 
\begin{equation}\label{eqn:betanew}
\tilde{\vbeta}^{(t)} =  \vbeta^{(t)} + u^{-1}\vX^{T}[\vy - b'(\vX\vbeta^{(t)})],
\end{equation}
which corresponds to a zero loss in the objective function. Thus, the constrained minimum of (\ref{eq:IHT-2}) is achieved by choosing the $k$ largest (in absolute value) components of $\tilde{\vbeta}^{(t)}$.
Consequently, $\vbeta^{(t+1)}$ in ($\ref{eq:IHT}$) has an explicit expression
\begin{equation}\label{eqn:Hk}
\vbeta^{(t+1)} = H_{k}[\tilde{\vbeta}^{(t)} ], 
\end{equation}
where $H_{k}[\vbeta]$
is the hard-thresholding operator setting all but the $k$ largest  components in $|\vbeta|$ to zero.

\begin{algorithm}[t]

\caption{SMLE-screening via IHT}\label{alg:IHT}

\begin{algorithmic} 
\REQUIRE Data $(\vy,\vX)$, screening size $k$, initial $\vbeta^{(0)}$, initial $u^{-1}_{0}$ and decrease rate $\tau \in (0,1)$
\vspace{0.1cm}

\noindent set $t = 0$ \\
\vspace{0.1cm}

\textbf{repeat} until stopping criterion is satisfied \{  \\

\quad set $\nu=u^{-1}_{0}$, $r=0$ \\

\vspace{0.1cm}

 \quad \textbf{repeat} until $l(\tilde\vbeta) \geq l(\vbeta^{(t)})$  \{ \\

 \quad \quad Compute (\ref{eqn:betanew}) and (\ref{eqn:Hk}): $\tilde\vbeta = H_k[\vbeta^{(t)} + \nu \vX^{T}[\vy -  b'(\vX\vbeta^{(t)})]]$ 
 

\quad \quad  $\nu = \tau \nu$ \\
\quad \quad $r =r + 1$ 

\quad \}  \\ 

\quad $\vbeta^{(t+1)} =\tilde\vbeta$ \\
\quad u-search$^{(t)}=r$\\
\quad $t = t + 1$  \\

\}  

\textbf{Output:} $\vbeta^{(t)}$, the number of iterations $t$, the number of $u$-search tries in each iteration, and an index set of retained features $\hat{s} = \{1\leq j \leq p :  \beta^{(t)}_{j} \neq 0\}$

\end{algorithmic}

\end{algorithm}

In IHT, $u^{-1}$ serves as a step size, which controls the distance moved from $\vbeta^{(t)}$ to $\vbeta^{(t+1)}$. While a larger $u^{-1}$ often helps to boost the iterations, the procedure may fail to converge when $u^{-1}$ is overly large. Thus, to balance algorithm convergence and iteration efficiency, one may choose to adaptively tune $u^{-1}$ at each step (called $u$-search). \cite{Chen+Chen:2014} proved that, when $u^{-1}$ is small enough, the IHT procedure leads to a non-decreasing likelihood. In our package, we first initialize $u^{-1}$ with a large value and then adaptively decrease its size  by a factor of $\tau \in (0,1)$ until $l(\vbeta^{(t+1)}) \geq l(\vbeta^{(t)})$ is satisfied. This seems to be an effective way for achieving sufficiently fast convergence.

We summarize the SMLE-screening procedure via IHT in Algorithm \ref{alg:IHT}. As can be seen from the algorithm summary, the procedure involves only simple numerical operations and adaptively tuning $u^{-1}$ often requires only a few tries to succeed. At each iteration, the joint information carried in $\vX$ is
naturally accounted for as a basis for the next update. These merits make SMLE-screening attractive in ultrahigh-dimensional data analysis, where computational hurdles and complex data structures are often faced.

\cite{Chen+Chen:2014} showed that, with appropriate $u^{-1}$ and $\vbeta^{(0)}$, the IHT updates lead to a local maximum of problem (\ref{eq:SMLE}), which provides an index set of $k$ important features, $\hat{s}$, corresponding to the non-zero entries of $\vbeta^{(t)}$. Based on $\hat{s}$, we then obtain a refined feature set from $\vX$ for subsequent in-depth model fitting. Under some regularity conditions, $\hat{s}$ contains all the relevant features with probability tending to one even when $p\gg n$, and thus is consistent for feature screening (sure screening). 

The IHT updates can be viewed as a member of the Majorize-Minimization algorithms. Its practical performance is affected by a series of implementation decisions such as the choice of initial value, stopping criterion, screening size $k$, and $u$-search. We address those algorithm details in Section \ref{subsec:main func}.

 \begin{algorithm}[t]
\caption{Post-screening selection with SMLE}\label{alg:smle_select}
\begin{algorithmic} 
\REQUIRE  Data  $(\vy,\vX_s)$, selection criterion,  sparsity lower and upper bounds $k_{min}, k_{max}$.\\

\vspace{0.1cm}

\textbf{for} $k$ \textbf{from} $k_{min}$ \textbf{to} $k_{max}$ \{   \\
\quad Obtain a sub-model $s_k$ by running Algorithm 1 with sparsity $k$ on $(\vy,\vX_s)$ \\

\quad Evaluate $s_k$ by computing a score $C_k$ based on the input selection criterion \\
\} \\
\vspace{0.1cm}

\textbf{Output} sub-model $s^{*} \in \{s_{k_{min}}, \ldots, s_{k_{max}}\}$ with the smallest evaluation score $C_k$

\end{algorithmic}
\end{algorithm}

\subsection{Post-screening selection with SMLE} \label{subsec:post}

In \cite{Chen+Chen:2014}, SMLE is mainly proposed for feature screening, the goal of which is to remove most irrelevant features before a more elaborate analysis. In practice, it is very likely that the feature set retained after screening still contains some irrelevant features. In principle, users can apply any well-developed selection method on the retained feature set to further identify relevant features.

In particular, one may further use the idea of SMLE to conduct post-screening selection. Specifically, assume that SMLE-screening was done and $q$ features were retained; we obtain a refined $n \times q$ feature matrix $\vX_s$. When $q$ is moderate, we can conveniently obtain a series of sub-models by running Algorithm 1 on $(\vy, \vX_s)$ with sparsity $k$ varying from $k_{{min}} \geq 1$ to  $k_{{max}} \leq q$. A final sub-model can then be selected based an information criterion such as AIC \citep{AIC}, BIC \citep{BIC}, and EBIC  \citep{Chen+Chen:2008}. 
We summarize this post-screening selection method in Algorithm 2, which inherits all the numerical merits from Algorithm \ref{alg:IHT}. In particular, the joint information in $\vX_s$ is naturally accounted for in the selection process. 
  
Technically, Algorithm \ref{alg:smle_select} can be used directly on $\vX$ without the need for feature screening. Its performance is actually quite impressive in our simulation studies. Nevertheless, when $p$ is very large, we do recommend using Algorithm \ref{alg:smle_select} only after Algorithm \ref{alg:IHT} for improved accuracy and stability.

\section[Implementation details]{Implementation details} \label{sec:imple}

The \proglang{R} package \pkg{SMLE} provides a set of functions for ultrahigh-dimensional feature screening under generalized linear models.
\pkg{SMLE} can be installed from the Comprehensive \proglang{R} Archive Network (CRAN) at \url{https://cran.r-project.org/web/packages/SMLE} using the following commands: 
\begin{Code}
install.packages("SMLE")
library(SMLE)
\end{Code}
\pkg{SMLE} requires the \proglang{R} packages 
\pkg{glmnet} \citep{Friedman2010,Rglmnet} for parameter initialization,  \pkg{mvnfast} \citep{mvnfast} and \pkg{matrixcalc} \citep{matrixcalc} for simulating correlated data. 

In this section, we first briefly describe the anticipated work flow for a data analysis using \pkg{SMLE}. We then provide a brief overview of the main functions of \pkg{SMLE} and illustrate the use and output of the functions through simple function calls. Detailed numerical examples are provided in Section \ref{sec:simu}.

\subsection{SMLE work flow}

\begin{figure}[t]
\centering
\includegraphics[width=12cm]{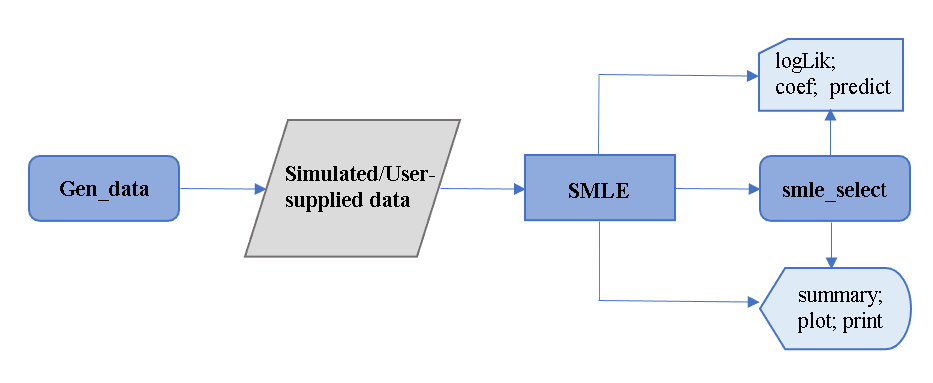}
\caption{Flowchart for calling functions in the \pkg{SMLE} package.}\label{fig:flowchart}
\end{figure}

The main \pkg{SMLE} functions are listed in Table \ref{tab:functions}. The work flow for a typical data analysis with \pkg{SMLE} is illustrated in Figure \ref{fig:flowchart} and is summarized as follows: 

\begin{table}[tb]
\centering
\begin{tabular}{p{0.25\textwidth}p{0.65\textwidth}}
\hline
Function Name          &   Description \\ \hline
\fct{Gen\_Data}        &   Simulate ultra-high dimensional GLM data \\
\fct{SMLE}             &   Joint feature screening using the SMLE method\\
\fct{smle\_select}     &   Post-screening feature selection\\
\fct{predict}    &   Fitted or predicted values under the final model \\ 
\fct{plot}             &   Plot method to evaluate SMLE screening/selection  for objects of class \class{smle} or \class{selection}\\ \hline
\end{tabular}
\caption{\label{tab:functions} Main \pkg{SMLE} functions and their brief descriptions.}
\end{table}

\begin{itemize}
\item For simulation studies or for testing purposes, the function \fct{Gen\_Data} can be used to simulate data assuming the response variable follows a GLM that depends on a small subset of possibly-correlated features. \fct{Gen\_Data} returns an object of class \class{sdata} that contains the matrix of features, the response variable, and additional information about the model used to simulate the data.

\item The function \fct{SMLE} is the main function based on Algorithm \ref{alg:IHT} for screening out features that are unlikely to be related to the response variable. Users can pass a \class{sdata} object, a data frame, or a matrix containing the data to the function. \fct{SMLE} returns a \class{smle} object that contains the top $k$ features and additional information about the screening process.

\item The function \fct{smle\_select} is used to conduct accurate post-screening feature selection based on Algorithm \ref{alg:smle_select}. Users can choose to implement this function with a selection criterion such as AIC, BIC or EBIC. Users can pass either a \class{smle} object or a data frame (matrix) to the function. \fct{smle\_select} returns an object of class \class{selection} that provides information about the selected features and the selection process. Users may also choose to run \fct{smle\_select} as a built-in option within the main function \fct{SMLE} by setting the argument \code{selection = TRUE}.

\item The \fct{plot} function for a \class{smle} or \class{selection} object can be used to visualize the screening or selection results.

\item The \fct{predict} function returns the fitted or predicted response values based on the features retained in a \class{smle} or \class{selection} object.

\item The package also includes several functions with usage similar to existing \proglang{R} functions for summarizing objects and fitting regression models (e.g. \fct{summary}, \fct{coef}, \fct{logLik}). 
\end{itemize}

\subsection{Main functions and arguments}\label{subsec:main func}

\subsubsection{Simulating ultrahigh-dimensional GLM data}\label{sssec:gendata}

The motivation for the function \fct{Gen\_Data} is to provide a freely available tool for simulating ultrahigh-dimensional GLM datasets with complex correlation structures between features. Some of the correlation 
structures available in this function were used in \citet{Chen+Chen:2014} to compare feature screening approaches. 
Both numerical and categorical features are permitted, and the number of features can be larger than the sample size. Users can choose the sample size $n$ and the total number of features $p$ in the dataset. 

Numerical features are sampled from a normal distribution with one of four commonly-used correlation structures. The strength of correlation is controlled by a parameter $\rho$, which can optionally be set by the user using the argument \code{correlation}. The four different correlation structures are:
\begin{description}
\item [Independence (ID) ] All features are independently sampled from a standard normal distribution. 
\item [Moving average (MA) ] Features are jointly normal with covariance (correlation) $\rho$ between adjacent features, $\rho/2$ between features two indices apart, and 0 otherwise. For example, the covariance matrix for four features would be:
$$
\left [
\begin{array}{cccc}
1&\rho&\rho/2&0\\
\rho&1&\rho&\rho/2\\
\rho/2&\rho&1&\rho\\
0&\rho/2&\rho&1
\end{array} \right ]
$$
\item [Compound symmetry (CS)] Features are jointly normal with 
covariance $\rho/2$ if both features are causally related (relevant) to the response variable, and with covariance $\rho$ otherwise. For example, with four features and assuming that features one and four are relevant, the covariance matrix used to simulate the features would be
%
%
$$
\left [
\begin{array}{cccc}
1&\rho&\rho&\rho/2\\
\rho&1&\rho&\rho\\
\rho&\rho&1&\rho\\
\rho/2&\rho&\rho&1
\end{array} \right ]
$$
\item [Auto-regressive (AR)] Features are jointly normal with covariance ${\rm cov} (x_j,x_h) = \rho ^ {|j-h|}$ for the $j$th and $h$th features with $j, h \in \{1, \ldots, p\}$. 
\end{description}

Categorical features are generated by first binning a numerical feature that was simulated as described above. The number of groups (levels) for a categorical feature is specified with \code{level_ctgidx}. After binning, the feature is converted from class \class{numeric} to \class{factor} and each bin is assigned a character from `A' to `Z'. Users are able to specify the number of categorical features in the dataset with the argument \code{num\_ctgidx}, and their positions in the dataset with the argument \code{pos\_ctgidx}.

 The response variable is simulated by assuming a GLM and that only a subset of the features are influential on the response. Normal, binary, and Poisson response variables are all available; the model is chosen with the argument \code{family}, as with the \proglang{R} function \fct{glm}. The user can choose the number of influential features with the argument \code{num\_truecoef}, and optionally, their positions in the feature matrix and their model effects with the arguments \code{pos\_truecoef} and \code{effect_truecoef}, respectively. If the positions of the influential features are not provided, they would be chosen randomly.

\fct{Gen\_Data} returns an object of class \class{sdata} containing the response vector $\vy$,  the $n \times p$ feature matrix $\vX$, and the coefficients for the features affecting the response.

The following code shows how to simulate a dataset with $n=200$ observations and $p=1000$ features, the first three of which are categorical, with three, four, and five levels, respectively. The response variable is generated based on a normal linear model with five influential features chosen by the function default.

\begin{Sinput}
R> set.seed(1)
R> Data_ctg <- Gen_Data(n = 200, p = 1000, family = 
"gaussian", pos_ctgidx = c(1,2,3), level_ctgidx = c(3,4,5))
R> head(Data_ctg$X)[,1:5]
\end{Sinput}

\begin{Code}
  C1 C2 C3          N4         N5
1  A  B  C -1.29171904  0.1370216
2  A  B  D  0.90967045 -1.2996452
3  B  B  A -1.10775568  1.1514089
4  B  C  C -0.38412387  1.5134475
5  B  C  D  0.08273483  0.8021379
6  B  A  D -0.48388247 -0.4305369
\end{Code}

\subsubsection{Joint feature screening}\label{sssec:smle}

\begin{figure}[t]
\centering
\includegraphics{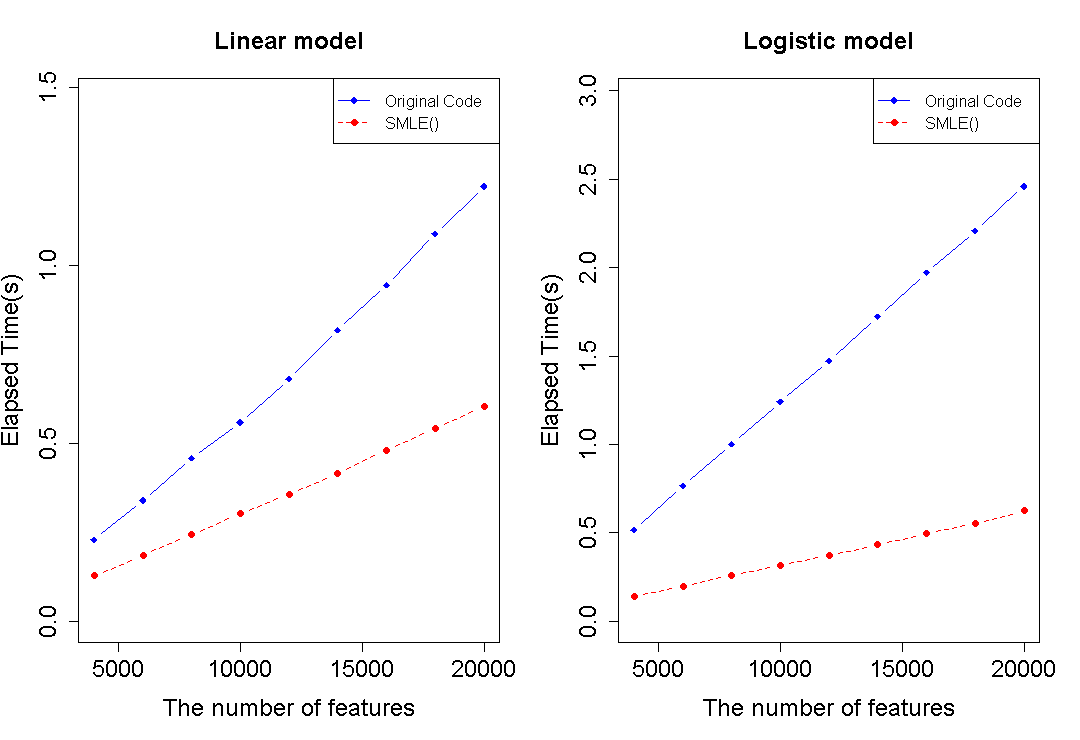}
\caption{Running time comparison between \fct{SMLE} and the original IHT implementation in \cite{Chen+Chen:2014}.}\label{fig:time}
\end{figure}

The main goal is to identify a manageable set of $k < p$ features that are  most related to the response variable. To that end, \fct{SMLE} is used to screen out features unlikely to be influential (i.e. irrelevant features); it serves as a pre-processing step before an in-depth analysis. Users can pass information about the input data $(\vy,\vX)$ to \fct{SMLE} via a \class{sdata} object, a data frame, or data matrices. When data are not input from a \class{sdata} object, the user should further specify the type of GLM with the argument \code{family}. The function conducts effective feature screening based on Algorithm \ref{alg:IHT}, which naturally incorporates the joint effects among features. In \fct{SMLE}, we make use of the \proglang{R} function \fct{crossprod} to handle the ultrahigh dimensional matrix product involved in Algorithm \ref{alg:IHT} without doing matrix transpose; this leads to improved computational efficiency in comparison with the original implementation in \cite{Chen+Chen:2014}. In Figure \ref{fig:time}, we show the elapsed time  \footnote{This example is conducted using the same computing platform described in Section 4. The same stopping rule is used for both implementations. The running time is averaged based on 100 repetitions.} (in seconds) of \fct{SMLE} and the original IHT code on a series of datasets generated by \code{Gen\_Data(correlation = "ID")} with $n=200$ and $p$ varying from 4000 to 20000.


In Table \ref{tab:smle}, we list the main arguments of \fct{SMLE} for specialized users to control the screening process. In particular, the argument \code{k} controls the number of important features to be retained from $\vX$ after screening. The choice of  \code{k} should reflect a user's belief or prior knowledge on the total number of relevant features for the input data. Intuitively, a larger \code{k} increases the chance of retaining all relevant features, while a smaller \code{k} brings more interpretive value and computational convenience for the subsequent in-depth analysis. One practical strategy is to set \code{k} to be three times larger than the anticipated number of relevant features.  In \fct{SMLE}, the default value of \code{k} is the largest integer not exceeding $0.5 \log(n) n^{1/3}$, which is recommended in \cite{Chen+Chen:2014} from a theoretical perspective. 

As mentioned in Section \ref{subsec:smle}, the choice of initial value $\beta^{(0)}$ has a non-negligible impact on SMLE-screening, as Algorithm \ref{alg:IHT} only leads to a local maximum. Since a non-decreasing likelihood is obtained by IHT updates, technically $\vbeta^{(t)}$ from Algorithm \ref{alg:IHT} can be always viewed as an improvement of $\vbeta^{(0)}$ in terms of the joint likelihood. In particular, when SMLE is used on linear models with $\vbeta^{(0)}=0$, the screening result based on $\vbeta^{(1)}$ coincides with the SIS-screening based on marginal effects; $\vbeta^{(2)}$ further improves SIS by incorporating joint information in $\vX$. This suggests that setting $\vbeta^{(0)}=0$ may be a reasonable choice in many applications. In \fct{SMLE}, users can specify $\vbeta^{(0)}$ with the argument \code{Coef_initial} based on their prior knowledge on the model coefficients. The default value of \code{Coef_initial} is the Lasso \citep{Tibshirani1996} estimate with the largest sparsity not exceeding $n-1$, which is implemented by the function \code{glmnet(pmax = n-1)} in the \proglang{R} package \pkg{glmnet}. This empirical choice seems to work well in our numerical studies.

In \fct{SMLE}, the initial value for $u^{-1}$ required in Algorithm \ref{alg:IHT} is set to be $1/\|\vX\|^2_{\infty}$, where $\|.\|_{\infty}$ is the matrix infinity norm denoting the maximum absolute row sum of a matrix. The use of $\|\vX\|^2_{\infty}$
serves as a computationally convenient replacement for computing the largest eigenvalue of
$\vX^T \vX$, which is used in \cite{Chen+Chen:2014} as a theoretical guidance for choosing $u$. Users can specify the decreasing rate $\tau \in (0,1)$ in $u$-search with the argument \code{U_rate}, whose default value is 0.5.  

\fct{SMLE} terminates the IHT iterations when $\|\vbeta^{(t)}- \vbeta^{(t-1)}\|_{2}$ is below the tolerance level specified in the argument \code{tol}. Since our goal here is feature screening, a large number of iterations $t$ may not always be necessary. We observe that, in many of our numerical examples, SMLE can successfully identify all the relevant features within a few iterations by estimating the corresponding coefficients to be non-zero. Therefore, when an accurate coefficient estimate is not needed, users may choose to set the argument \code{fast = TRUE}, which allows early stopping of Algorithm \ref{alg:IHT}. Specifically, with \code{fast = TRUE}, \fct{SMLE} terminates the IHT iterations when one of the following rules is satisfied:
\begin{itemize}
    \item $\|\vbeta^{(t)}-\vbeta^{(t-1)}\|_{2}$ $<$ \code{k}$^{1/2}$ $\times$ \code{tol} 

    \item $l(\vbeta^{(t)}) - l(\vbeta^{(t-1)}) < 0.01\ [\ l(\vbeta^{(1)}) - l(\vbeta^{(0)})\ ]$
    
   \item  The non-zero entries in $\{\vbeta^{(t)}\}$ remain unchanged for 10 consecutive iterations.
\end{itemize}

\begin{table}[t]
\centering
\begin{tabular}{p{0.15\textwidth}p{0.6\textwidth}p{0.15\textwidth}}
\hline
Arguments  &  Description & Default Value\\ \hline

\code{k}  & Total number of features to be retained after screening. & $\frac{1}{2} \log(n) n^{1/3}$ \\

\code{keyset}  &  A vector to indicate a set of key features that are forced to remain in the model.  & NULL\\

\code{Coef_initial} & Initial coefficient value $\vbeta^{(0)}$ for IHT. & \code{glmnet(pmax=n-1)} \\

\code{categorical} & Logical flag for whether the input feature matrix includes categorical features. & NULL \\

\code{group}	&  Logical flag for whether to treat the dummy variables corresponding to each categorical feature
as a group. & TRUE\\

\code{tol}  & A tolerance level for $\|\vbeta^{(t)} - \vbeta^{(t-1)}\|_{2}$ to stop the iteration. &  $10^{-2}$\\

\code{fast} &  Logical flag to enable early stop for IHT. & FALSE \\

\code{U\_rate} &  Decrease rate in $u$-search. &  0.5\\

\hline
\end{tabular}
\caption{Main arguments for the \fct{SMLE} function.}\label{tab:smle}
\end{table}

When the input data for \fct{SMLE} contains categorical features, users should set the argument \code{categorical = TRUE}, which enables the function to automatically detect the locations and levels of the \code{factor} columns in the feature matrix. By default, a $L$-level categorical feature would be encoded by $L-1$ dichotomous dummy variables, which are to be retained or screened out together if \code{group = TRUE}.

With the \code{keyset} argument, users can choose to manually keep a subset of features in the SMLE-screening process; this can be useful when some features are known to be influential or confounding. The following code demonstrates the usage of \fct{SMLE} with \code{keyset} on the dataset \code{Data_ctg} generated in the earlier part of this subsection. Here we ensure that the first (\code{C1}), the fourth (\code{N4}), and the fifth (\code{N5}), features are always kept during the IHT updates and are therefore surely retained after screening. In this example, since we choose to retain \code{k = 15} features in total, the retained set would include the 3 features specified in \code{keyset} plus 12 features to be suggested by Algorithm \ref{alg:IHT}. When \code{keyset} contains categorical features, it is required to have \code{group = TRUE}.

\begin{Sinput}  
R> fit  <-  SMLE(Y = Data_ctg$Y, X = Data_ctg$X, k = 15, family = "gaussian", 
keyset = c(1,4,5), categorical = TRUE, group = TRUE)
\end{Sinput}

\fct{SMLE} returns an object of class  \class{smle}, which contains information regarding the input data, model assumption, IHT updates, and the screening results.

\subsubsection{Post-screening selection}

As mentioned in Section \ref{subsec:post}, the features retained after screening are still likely to contain some that are not related to the response. The function \fct{smle\_select} is designed to implement Algorithm \ref{alg:smle_select} to further identify the relevant features.

\fct{smle\_select} can be applied to a \class{sdata} object, a \class{smle} object, or a user-supplied dataset $(\vy, \vX_s)$. As discussed before, when $p$ is very large, a screening step is usually needed before \fct{smle\_select} can be efficiently applied. 
In the package, we provide an option to automatically run \fct{smle\_select} after screening within the main function \fct{SMLE}.

\begin{table}
\centering

\begin{tabular}{p{0.2\textwidth}p{0.4\textwidth}p{0.3\textwidth}}
\hline
Arguments & Description &  Default Value \\ \hline
\code{k\_min} (\code{k\_max})  & The lower (upper) bound for candidate model sparsity.   & $\mbox{\code{k\_min}} = 1$, \code{k\_max} = number of input features\\

\code{sub\_model} &	 	
A index vector indicating which features are to be selected. Not applicable if a \class{smle} object is the input.  & NULL \\

\code{criterion}	   & Selection criterion to be used. One of \code{"ebic"},\code{"bic"},\code{"aic"}. &   \code{"ebic"}\\

\code{gamma\_ebic}	& The EBIC parameter in $[0 , 1]$.  &  0.5\\

\code{gamma\_seq} & The sequence of values for \code{gamma\_ebic} when \code{vote = TRUE}. &   \code{(0,0.2,0.4,0.6,0.8,1)}\\

\code{vote} &  The logical flag for whether to perform the voting procedure, when \code{criterion = "ebic"}. & FALSE  \\

\code{vote\_threshold} & A relative voting threshold in percentage. A feature is considered to be important when it receives votes passing the threshold. &  0.6\\

\code{para} &Logical flag to use parallel computing to do selection.& FALSE\\
\hline
\end{tabular}
\caption{Main arguments for the \fct{smle\_select} function.}\label{tab:smle_select}
\end{table}

We list the main arguments of  \fct{smle\_select} in Table \ref{tab:smle_select}. Users can set the range or specify a subset of features to be further selected. With the argument \code{criterion}, users can choose to run \fct{smle\_select} using their preferred selection criterion. 

When the criterion EBIC is used, users may further specify its tuning parameter using the argument \code{gamma\_ebic}. Alternatively, one may set \code{vote = TRUE}, which would repeat the EBIC-based selection with a sequence of tuning parameters provided in \code{gamma\_seq}. With different tuning parameters, different sets of features would be selected; a feature is considered to be important when it is selected more than \code{vote_threshold} of times in the entire procedure. Users may change this threshold to enlarge or reduce the size of the selected model by using the \fct{vote\_update} function, which is a simple method for a \class{selection} object that avoids the need to re-run \fct{smle\_select}.

The following code demonstrates the use of \fct{smle\_select} with EBIC voting on the \class{smle} object \code{fit} generated before. The output \code{fit_s} is an object of class \class{selection}, which contains the information regarding the selection procedure and result.

\begin{Sinput}
R> fit_s <- smle_select(fit, criterion = "ebic", gamma_seq = seq(0,1,0.2),
vote = TRUE)
\end{Sinput}

In \fct{smle\_select}, the selection is done by evaluating a sequence of sub-models with sizes varying from \code{k\_min} to \code{k\_max}. When the package is run in a multi-core computing environment, users may opt to use parallel computing to boost this selection procedure. Specifically, by setting \code{para = TRUE}, the creation and evaluation of the sub-models would be distributed to multiple computing cores for parallel processing. The implementation of the parallel option depends on the user's operating system: for Unix and Mac users, \fct{mclapply} is used \citep{R}; for Windows users,  \fct{parLapply} is used \citep{R}. Note that \fct{smle\_select} is mainly designed for post-screening selection, where the number of candidate features is moderate and the associated computational cost is often acceptable. Considering the communication cost between cores, parallel processing may not necessarily lead to a significant speed-up for the regular usage of \fct{smle\_select}.

\begin{figure}[t]
\centering
\includegraphics{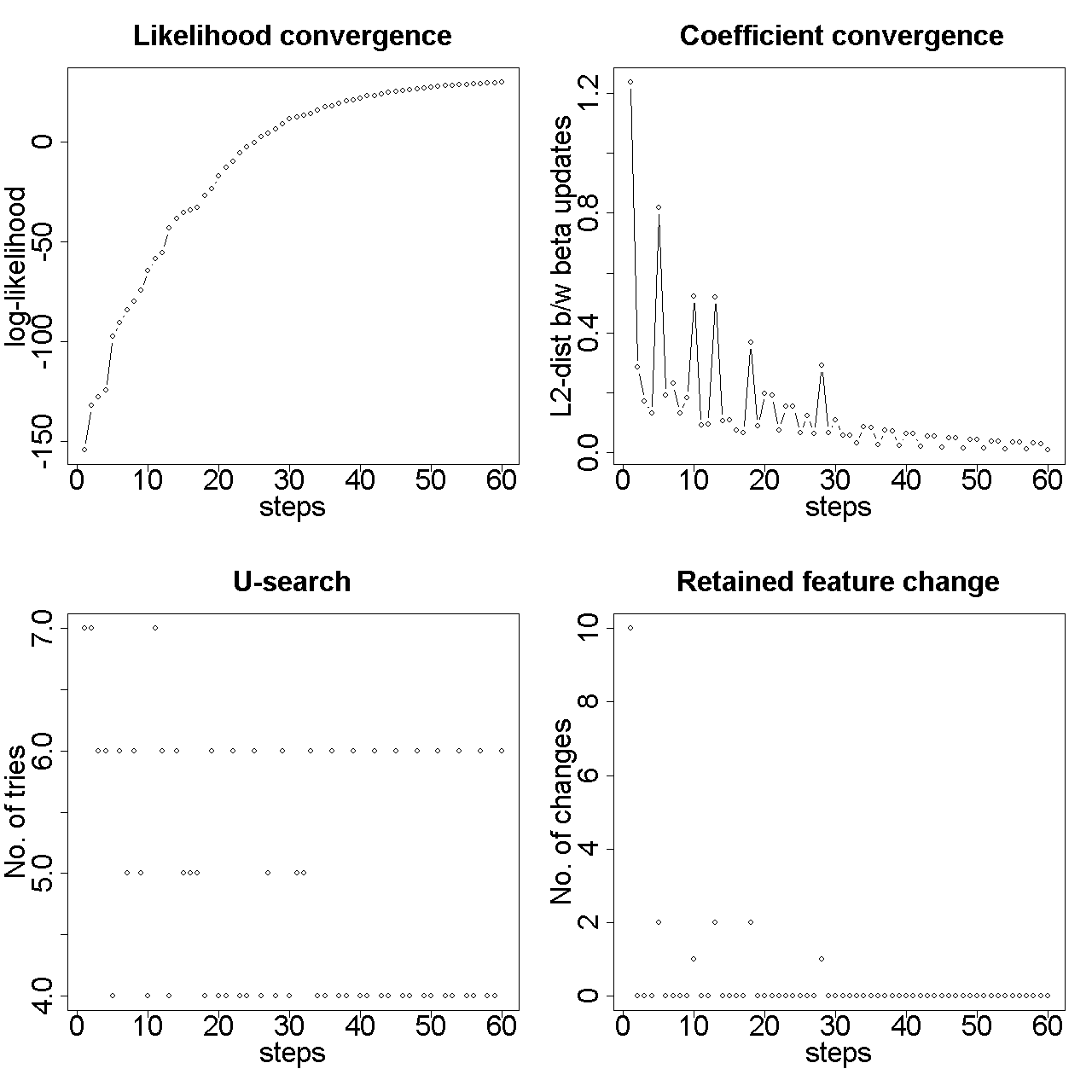}
\caption{Example plot for \class{smle} class - IHT Convergence with $\vbeta^{(0)}=0$.}  \label{fig:smleplot1} 
\end{figure}


\subsubsection{Plotting}

Plot functions have been included for both the \class{smle} and \class{selection} classes. The \fct{plot} function for class \class{smle} returns two plot windows. By default, the first plot window contains four subplots each showing a measure of IHT convergence on the vertical axis and iteration index on the horizontal axis; see Figure  \ref{fig:smleplot1} as an example. The convergence measures are:
 \begin{itemize}
     \item log-likelihood (top left)
     \item Euclidean distance between the current 
 and previous coefficient estimates (top right)
 
     \item the number of $u$-search tries in Algorithm \ref{alg:IHT} (bottom left)
     
     \item the number of membership changes in the retained feature set between the current and the previous IHT updates (bottom right)
 \end{itemize}

\begin{figure}[t]
\centering
\includegraphics[width=12cm]{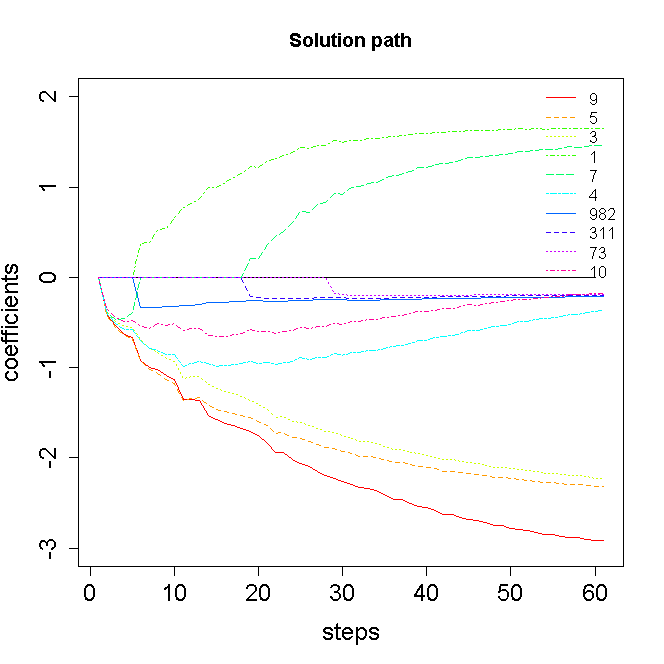}
\caption{Example plot for \class{smle} class - IHT solution path with $\vbeta^{(0)}=0$.} \label{fig:smleplot2}
\end{figure}


The second plot window shows the solution path (estimated coefficient by IHT iteration) for selected features; see Figure \ref{fig:smleplot2} as an example. This plot provides a direct insight on how a coefficient changes over the iterations. By default, the solution path is given for all the relevant features suggested in the input \class{smle} object. Users may choose to plot a solution path for a customized group of features. This can be done by plotting only the top \code{num\_path} features or plotting only the features specified in the argument \code{which\_path}.

Users can optionally change which figure appears in the second plot window. For example, passing the argument \code{outplot = 1} will cause the log-likelihood to appear in the second plot window; the solution path will instead be plotted in the first plot window. Any additional arguments passed to \fct{plot} will be used when creating the plot in the second plot window.

\begin{figure}[t]
\begin{subfigure}{.5\textwidth}
\centering
\includegraphics[width=6cm]{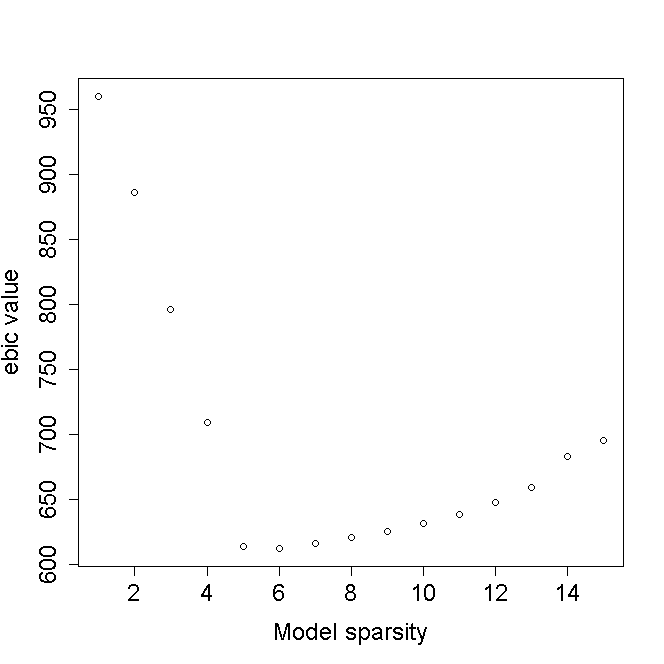}
 \caption{Selection criterion}
  \label{fig:sfig1}
\end{subfigure}%
\begin{subfigure}{.5\textwidth}
\centering
\includegraphics[width=6cm]{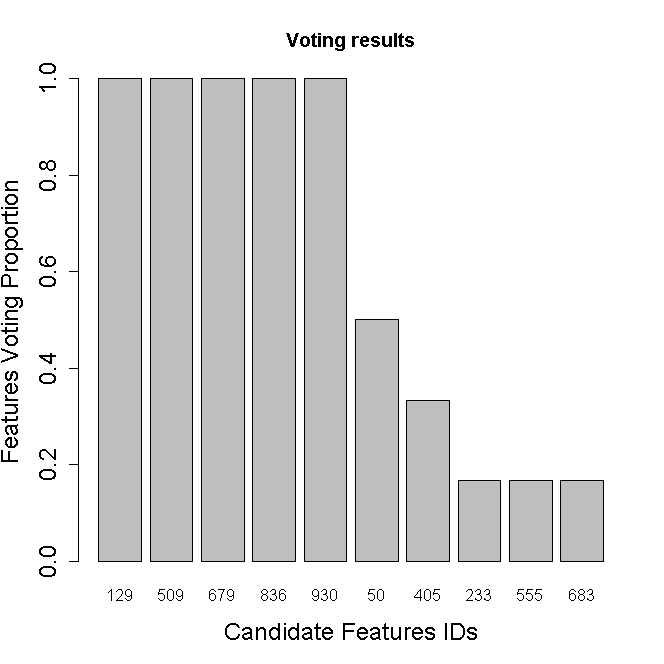}
 \caption{ Voting results}
  \label{fig:sfig2}
\end{subfigure}
\caption{\label{fig:selection}Example plot for \class{selection} class.}
\end{figure}

The \fct{plot} function for an object of \class{selection} class returns a plot showing the selection criterion scores evaluated for the candidate sub-models with varying sizes (number of features); see Figure \ref{fig:selection}(a) as an example. A selection curve typically has a "V" shape: the curve decreases at the beginning as sub-model size increases before it increases again when there is little benefit in including additional features. For example, Figure \ref{fig:selection}(a) suggests a sub-model with six features, which are to be treated as influential ones. When criterion EBIC is used with \code{vote = TRUE}, \fct{plot} additionally returns a voting plot showing the relative inclusion frequency of each feature in the sequence of sub-models generated by Algorithm \ref{alg:smle_select}; see Figure \ref{fig:selection}(b).

\subsubsection{Prediction}

The \fct{predict} function in \pkg{SMLE} is based on the generic \fct{predict} function in \pkg{stats}; it returns predicted response values based on a model containing only the features retained in a \class{smle} object or selected in a \class{selection} object. The predictions are made by fitting the model using the \fct{glm} function.

As with the \fct{predict} function for class \class{glm} or \class{lm}, users can optionally specify a data frame with new values for the features in the model using the \code{newdata} argument. If \code{newdata} is not provided, predictions will be given for the original training data; this leads to the fitted response values.

When using \fct{predict}, the type of prediction required must be specified using the \code{type} argument. The default is \code{type = link} which returns predictions on the scale of the linear predictor; \code{type = response} returns predictions on the scale of the response variable. In particular, with \code{type = response},  \fct{predict} returns the predicted mean responses for a linear or Poisson model and the predicted success probabilities for a logistic model.

\subsubsection{Additional functions for regression-based objects}

\pkg{SMLE} has some additional functions that can be used to extract information about the model fit. Specifically, the function \fct{logLik} first refits the screened (or selected) model using \fct{glm} and then returns the log-likelihood of the fitted model. The function \fct{coef} can be conveniently used to extract regression coefficients from a fitted \class{smle} or \class{selection} object. In addition, \fct{summary} is extended from the base function to take a \class{smle} or \class{selection} object and displays summary information about the fitted model object.

\section[Examples]{Examples} \label{sec:simu}

We demonstrate the usage and effectiveness of \pkg{SMLE} using on a series of simulation studies. Numerical experiments were conducted on a Linux server with 2.2GHz CPUs via Compute Canada.

\subsection{Demo code for SMLE-screening}
\label{sec:4.1}

We first show how to use \pkg{SMLE} to conduct feature screening and post-screening selection via a simulated example. To this end, we use \fct{Gen\_Data} to generate a synthetic dataset with $n=400$ observations and $p=1000$ features. We generate the feature matrix $\vX$ from a multivariate normal distribution with an auto-regressive structure, where the adjacent features have a high correlation of $\rho=0.9$. The response variable $Y$ is generated based on the following logistic model with success rate $\pi$ and linear predictor:
\begin{equation*}
 \mbox{logit}(\pi) = 2x_1 + 3x_3 - 3x_5 + 3x_7 - 4x_9. 
\end{equation*} 
In this setup, the feature matrix contains only five features that are causally-related to the response, as indicated in the model.  

This dataset is generated by the following code; the resulting simulated data are stored in \code{Data\_eg}, which is an object of class \class{sdata}. Users can use the \fct{print} function to show properties of the  simulated data.

\begin{Sinput}
R> set.seed(1)
R> Data_eg <- Gen_Data(n = 400, p = 1000, family = "binomial",
correlation = "AR", rho = 0.9, pos_truecoef = c(1,3,5,7,9),
effect_truecoef = c(2,3,-3,3,-4))
R> print(Data_eg)
\end{Sinput}
\begin{Code}
Call:
 Gen_Data(n = 400, p = 1000, pos_truecoef = c(1, 3, 5, 7, 9), 
    effect_truecoef = c(2, 3, -3, 3, -4), correlation = "AR", 
    rho = 0.9, family = "binomial")
 
An object of class sdata
 
Simulated Dataset Properties:
 Length of response: 400
 Dim of features: 400 x 1000
 Correlation: auto regressive
 Rho: 0.9
 Index of Causal Features: 1, 3, 5, 7, 9
 Model Type: binomial
\end{Code}

We then run \fct{SMLE} to conduct SMLE-screening on the data example \code{Data\_eg}, assuming that the identities of the causal features were unknown; the goal is to obtain a refined feature set by removing most irrelevant features from $\vX$. The following code shows the simplest function call to \fct{SMLE}, where we aim to retain only $k=10$ important features out of $p=1000$.

\begin{Sinput}
R> fit1 <- SMLE(Y = Data_eg$Y, X = Data_eg$X, k = 10, family = "binomial")
R> summary(fit1)
\end{Sinput}
\begin{Code}

Call:
  SMLE(X = Data_eg$X, Y = Data_eg$Y, k = 10, family = "binomial")
 
An object of class summary.smle
 
Summary:

  Length of response: 400
  Dim of features: 400 x 1000
  Model type: binomial
  Retained model size: 10
  Retained features: 1, 3, 5, 7, 9, 68, 430, 536, 661, 709
  Coefficients estimated by IHT: 1.572, 3.127, -1.893, 2.534,
  -4.499, -0.348, -0.316, 0.579, 0.717, 0.444
  Number of IHT iteration steps: 59                                                                                                        
\end{Code}

The function returns a \class{smle} object in the variable called \code{fit1}. The \fct{summary} function confirms that a refined set of 10 features is selected after 59 IHT iterations. We can see that all 5 causal features used to generate the response are retained in the refined set. This indicates that screening is successful;  the dimensionality of the feature space  is reduced from $p=1000$ down to $k=10$ without losing any important information.

In the code that follows, we fit a marginal regression between the response and the second feature, $x_2$, in \code{Data\_eg}. From the true model, we know that $x_2$ is not causally-related to the response. Yet, we can see that the marginal effect of $x_2$ appears to be pretty high; thus, this irrelevant feature is likely to be retained in the model if the screening is done based on marginal effects only. In this example, \fct{SMLE} accurately removes $x_2$, as its screening naturally incorporates the joint effects among features.

\begin{Sinput}

R> coef(summary(glm(Data_eg$Y ~ Data_eg$X[,2], family = "binomial")))
\end{Sinput}

\begin{Code}
                  Estimate Std. Error   z value     Pr(>|z|)
(Intercept)     0.02440072  0.1125979 0.2167067 8.284369e-01
Data_eg$X[, 2] 1.10465766  0.1337063 8.2618250 1.434619e-16
\end{Code}

Note that the refined set returned in \code{fit1} still contains some irrelevant features; this is to be expected, as the goal of feature screening is merely to remove most irrelevant features before conducting an in-depth analysis. As discussed in Section \ref{subsec:post}, one may conduct an elaborate selection on the refined set to further identify the causal features.
In particular, the \fct{smle\_select} function in \pkg{SMLE} can be readily used for the purpose of post-screening selection. The following code shows how this function can be used on \code{fit1} with the selection criterion EBIC. As can be seen below, \fct{smle\_select} returns a \class{selection} object \code{fit1\_s}, which exactly identifies the five features in the true data generating model. 

\begin{Sinput}
R> fit1_s <- smle_select(fit1, criterion = "ebic")
R> summary(fit1_s)
\end{Sinput}
\begin{Code}
Call:
  smle_select(object = fit1, criterion = "ebic")
 
An object of class summary.selection
 
Summary:
 
  Length of response: 400
  Dim of features: 400 x 1000
  Model type: binomial
  Selected model size: 5
  Selected features: 1, 3, 5, 7, 9
  Selection criterion: ebic
  Gamma for ebic: 0.5
\end{Code}

As mentioned in Section \ref{subsec:main func}, users can visualize the above screening and selection results by using the \fct{plot} function.

\subsection{Screening performance}

Next, we test the performance of \pkg{SMLE} in terms of screening accuracy and efficiency. To this end, we use \fct{Gen\_data} to generate 500 independent datasets with $p=1000$ and $n= 100, 200,$ and $400$ from linear, Poisson, and logistic models, respectively. For all these datasets, a compound symmetry structure with $\rho=0.3$ is used, where the first four features (i.e. $x_1$, $x_2$, $x_3$, $x_4$) are used to generate the response variable. An equal coefficient is assigned to each of the four causal features; the value of the coefficient is set to 2.5,  0.7, and 1.5 for the three aforementioned models, respectively.

We use \fct{SMLE} in our package to conduct feature screening on these simulated datasets to retain only $k= 20, 10, 30$ important features for linear, Poisson, and logistic models, respectively. We measure the screening accuracy by sure screening rate (SSR) and
positive retaining rate (PRR). 
SSR is reported as the proportion of times that all causal features are retained after screening; PRR is calculated by the averaged proportion of causal features that are retained after screening. An averaged elapsed time (in seconds) for \fct{SMLE} in each model setup is reported as a measure of screening efficiency. We repeat the experiments with both \code{fast = FALSE} and \code{fast = TRUE}. For comparison, the screening performance of (I)SIS and  Lasso are also reported under the same setup. Following the recommendations from the corresponding packages, we conduct SIS by \code{SIS(iter = F)} and ISIS by \code{SIS(iter = T)} in package \pkg{SIS}, and conduct Lasso by \fct{glmnet} in package \pkg{glmnet}.

We summarize the simulation results in Table \ref{tab:comparision}, where all decimal numbers are truncated to have 2 digits. It can be seen that 
\fct{SMLE} has the highest accuracy for all three model setups in terms of both SSR and PRR; the other methods (packages) are heavily affected by the correlation among features. By the nature of Algorithm \ref{alg:IHT}, the high accuracy of \fct{SMLE} comes with a computational cost, but this is moderate in most cases. Considering the improved accuracy of \fct{SMLE}, this small computational investment seems to be quite worthwhile. The \code{fast = TRUE} option helps to improve the run time of \fct{SMLE} with a little sacrifice to the screening accuracy. Notably, even with \code{fast = TRUE}, \fct{SMLE} still achieves a promising accuracy that seems to be superior to the other two methods.

\begin{table}[t]
\centering

\begin{tabular}{lcccr}
\hline
Model& Functions & PRR & SSR & Time \\ \hline
\multirow{5}{*}{Linear}& \code{SIS(iter = F)}& 0.32&0.00&0.02\\&\code{SIS(iter = T)}&0.89&0.83&1.92\\&\code{glmnet()}&0.57&0.18&<0.01\\&\code{SMLE()}&1.00&1.00&0.21\\&\code{SMLE(fast = T)}&0.99&0.99&0.19\\ \hline

\multirow{5}{*}{Poisson}&\code{SIS(iter = F)}& 0.11&0.00&0.06\\&\code{SIS(iter = T)}&0.58&0.46&4.62\\&\code{glmnet()}&0.12&0.00&0.01\\&\code{SMLE()}&0.97&0.93&0.61\\&\code{SMLE(fast = T)}&0.96   &0.90&0.53\\ \hline

\multirow{5}{*}{Logistic}&\code{SIS(iter = F)}& 0.54&0.03&0.16\\&\code{SIS(iter = T)}&0.79& 0.62&9.27\\&\code{glmnet()}&0.84&0.54&0.04\\&\code{SMLE()}&0.97& 0.89&0.99\\&\code{SMLE(fast = T)}&0.96& 0.88  &0.86\\ \hline
\end{tabular}
\caption{\label{tab:comparision}Screening performance of \pkg{SMLE}.}
\end{table}

\subsection{Application to high-dimensional genetic data}

To further demonstrate \pkg{SMLE}, we applied it to a simulated high-dimensional genetic  dataset to detect associations between single-nucleotide polymorphisms (SNPs) and a response variable.
To get a realistic genetic dataset, the genotypes were sampled from genotypic distributions derived from the  1000 Genomes project, Phase 1 \citep{1000Genomes} using the \proglang{R} package \pkg{sim1000G} \citep{sim1000g}. 
Since the SNP distributions are derived from individuals from existing human populations, this dataset naturally has a complex correlation structure. In particular, it is believed that strong non-linear correlations are present between SNPs in close proximity from the same chromosome.

The dataset consists of $p=10,031$ SNPs from chromomes 14 through 22 on $n=800$ individuals. The genotypes were coded as 0, 1, or  2 by counting  the number of minor alleles (the allele that is less common in the sample). The continuous response variable was simulated from a normal distribution with mean that depends additively on the causal SNPs and a standard deviation of 8. The linear predictor for the response follows a polygenic model, which specified 31 SNPs with small to large effects. The goal here is to locate regions containing the 31 causal SNPs; the effects of those SNPs were determined as follows.

\begin{itemize}
    
    \item Twenty SNPs were randomly selected to have small effects, which were drawn from a $N(0, 0.1)$ distribution.
    
    \item Five SNPs were randomly selected to have moderate effects, which were drawn from a  $N(0, 0.5)$ distribution. 
    
    \item Four SNPs (labelled: rs2967291, rs1534941, rs112915930, rs6132052) were selected to have large magnitude effects. Their effects were fixed at 2,-2,1 and -1, respectively.
    
    \item Two SNPs were selected to have an effect on the response only through an interaction; no marginal effect was simulated. The two SNPs are labelled rs12608528 and rs11157211 and their interaction coefficient was set to be 4.
    
    \item The intercept was arbitrarily chosen to be 40 in order to avoid negative response values. 

\end{itemize}

 This simulated genetic dataset is included in \pkg{SMLE} and can be loaded with the following command.
\begin{Sinput}
R> data("synSNP")
\end{Sinput}
We save the response as a vector \code{Y_SNP} and the features as a matrix \code{X_SNP}.
\begin{Sinput}
R> Y_SNP <- as.matrix(synSNP[,1], ncol = 1)
R> X_SNP <- as.matrix(synSNP[,-1])
\end{Sinput}

\begin{figure}[t!]
\centering
\includegraphics[width=\textwidth]{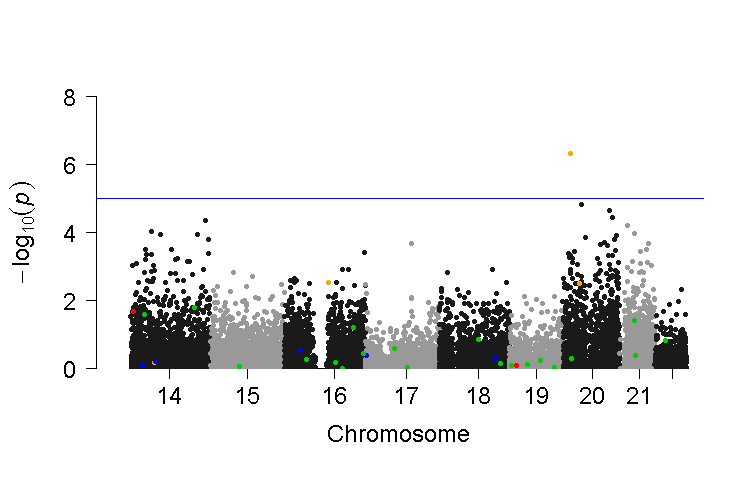}
\caption{\label{fig:mah} Manhattan plot of $-\log_{10}$(p-value) for the single-SNP association test. The blue line represents suggestive significance. The coloured points correspond to SNPs simulated to be causally related to the response in the order of importance: red, orange, blue, green. }
\end{figure}

The goal of study here is to flag genomic regions likely to contain the causal SNPs. This job can be typically done by identifying a small set of key SNPs and then marking the nearby locations. As a conventional approach, we first performed a single-SNP analysis using linear regression to test the marginal effects of the SNPs on the response. We report $-\log_{10}$ of the p-values of the corresponding estimated marginal coefficients using a Manhattan plot (Figure \ref{fig:mah}). The p-values of the 31 causal SNPs are coloured based on their effect sizes. The blue line shows the conventional threshold used to declare suggestive significance (p-value $< 10 ^{-5}$). We observed that only one SNP passed this threshold, which was simulated to have a large effect ("rs11157211"). All other SNPs did not show marginal significance and thus were not treated as the key ones. As a result, the conventional marginal-test-based approach failed to flag all the regions harboring the causal SNPs. 

We then ran \fct{SMLE} to flag genomic regions by retaining $k=40$ key SNPs after screening. The results were compared with those obtained by \fct{SIS}. The SMLE-screening and SIS-screening were carried out with the following code.

\begin{Sinput}
R> SMLE_fit <- SMLE(Y = Y_SNP, X = X_SNP, family = "gaussian", 
k = 40, fast = F)

R> SIS_fit <- SIS(y = Y_SNP, x = X_SNP, family = "gaussian", 
nsis = 40, iter = F)
\end{Sinput}
\begin{figure}[t!]
\begin{subfigure}{.5\textwidth}
\centering
\includegraphics[width=7.5cm]{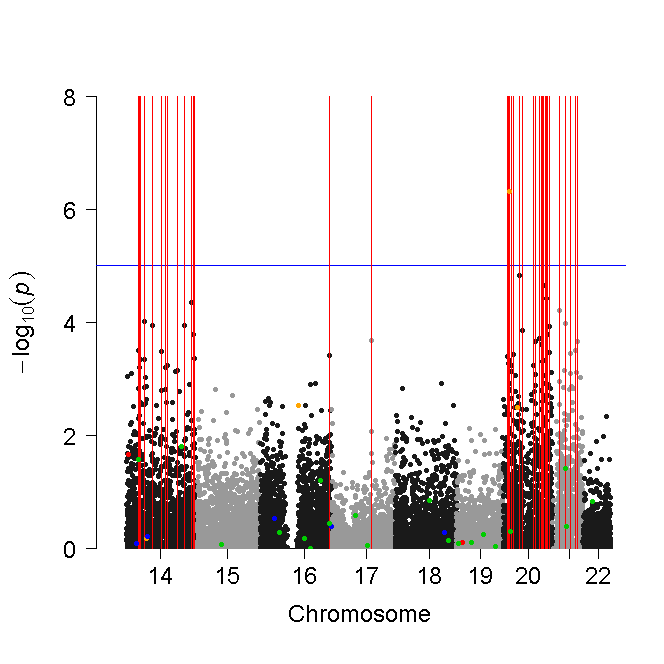}
\end{subfigure}%
\begin{subfigure}{.5\textwidth}
\centering
\includegraphics[width=7.5cm]{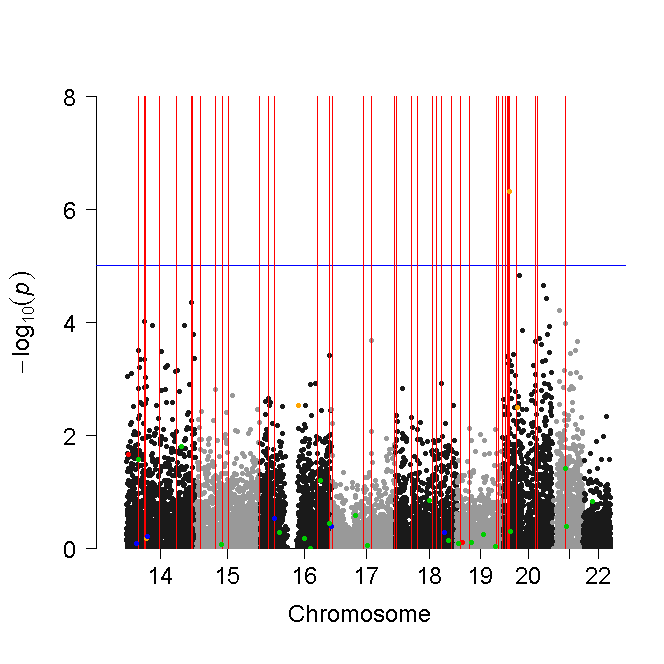}
\end{subfigure}
\caption{\label{fig:eff1} Important genetic regions flagged with red lines by \fct{SIS} (left) and \fct{SMLE}(right).}

\end{figure}
 In Figure \ref{fig:eff1}, we show the screening result from the code above. It is observed that SIS focused on chromosomes 14, 20 and 21, where SNPs have relatively larger marginal effects; in contrast, SMLE flagged regions across all chromosomes. 

To assess the screening performance, for each of the 31 causal SNPs we compute the minimum distance between that SNP and the SNPs in the retained set (MRD). A small MRD indicates that the retained set contains at least one SNP that is close to the causal SNP; consequently, genetic regions suggested by the retained SNPs are likely to contain the causal SNPs. Therefore, an effective screening is expected to have small MRDs for most causal SNPs.

\begin{figure}[t]
\begin{subfigure}{.5\textwidth}
\centering
\includegraphics[width=7.5cm]{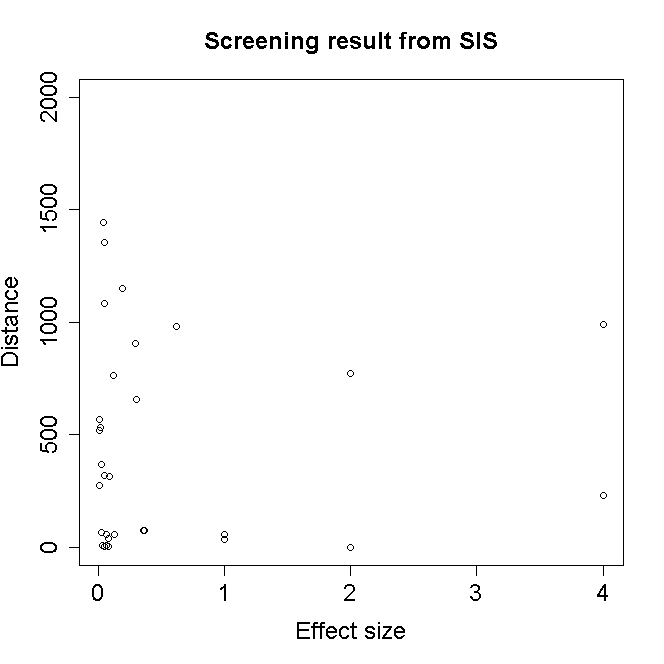}
\end{subfigure}
\begin{subfigure}{.5\textwidth}
\centering
\includegraphics[width=7.5cm]{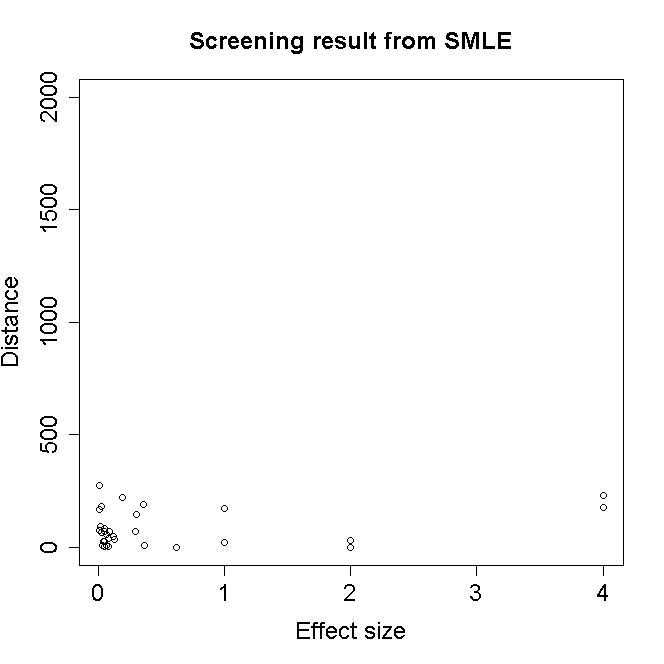}
\end{subfigure}
\caption{\label{fig:eff2}MRD on causal SNPs for \fct{SIS} (left) and \fct{SMLE} (right).}
\end{figure}

In Figure \ref{fig:eff2}, we show the MRDs of the 31 causal SNPs from the screening conducted with \fct{SMLE} (\code{SMLE_fit}) and \fct{SIS} (\code{SIS_fit}). The horizontal axis denotes the true effect sizes of the causal SNPs. Since \fct{SIS} only uses the marginal information, the 40 locations suggested by \code{SIS_fit} tend to be very close to the peak on chromosome 20 shown in Figure \ref{fig:mah}; this leads to large MRDs for most causal SNPs that are away from this peak. In comparison, \code{SMLE_fit} achieves a much smaller overall MRD. In particular, it effectively identifies the genetic regions containing the causal SNPs with the effect magnitude of 2 and shows a substantial improvement over \code{SIS\_fit} in identifying the two causal SNPs with a large interaction effect. 

\begin{figure}[t!]
\centering
\includegraphics[width=7.5cm]{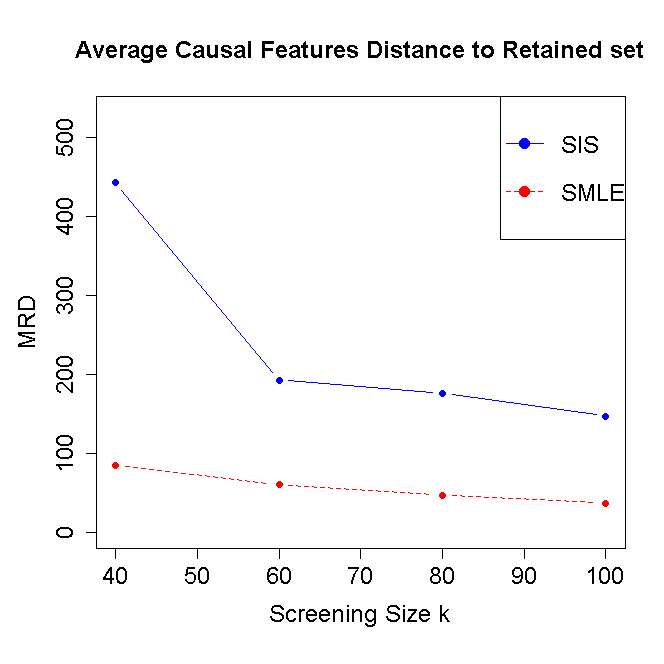}
\caption{\label{fig:eff3} The averaged MRD on causal SNPs for \fct{SIS} and \fct{SMLE} with varying $k$.}
\end{figure}

We repeated the analysis by increasing the model screening size, $k$. The corresponding averaged MRDs for all causal SNPs are shown in Figure \ref{fig:eff3}. The performance of both screening methods improve as $k$ increases. Compared with \fct{SIS}, \fct{SMLE} tends to be more stable; it conducts effective screening by consistently achieving a low averaged MRD
over different choices of $k$. The high reliability makes it a preferable choice in practice.

\section[Conclusion]{Concluding remarks}

In this paper, we introduced a user-friendly feature screening package \pkg{SMLE} as a powerful tool to process ultrahigh-dimensional GLMs. The package provides an efficient implementation of the joint screening method SMLE, which leads to more reliable screening results compared with the commonly-used marginal methods. In \pkg{SMLE}, users can also conduct accurate post-screening selection based on an accelerated IHT procedure with a preferred selection criterion. Plotting tools are provided to visualize the screening and selection results, which can be readily used for making inference or prediction.

We illustrated the usage of the main functions in \pkg{SMLE} with discussions and examples. The effectiveness and efficiency of the package are supported by extensive numerical studies.

The idea of SMLE has been demonstrated to be attractive in processing many types of ultrahigh-dimensional data. In the current version, the package focuses on feature screening for GLMs, which have been widely used in many scientific areas. It would be promising to extend the application scope of \pkg{SMLE} to other modeling scenarios such as Cox's models and multivariate response regression, where the SMLE-based methodology has been discussed in the literature. For future work, it is also of great interest to embed \pkg{SMLE} in a distributed framework, so that it may help to handle the big data situation, where high-dimensional data segments are stored in different places.

\section*{Acknowledgments}

This research was supported in part by NSERC grants RGPIN-2016-05024, RGPIN-2019-06051 and NSFC grant 11690014. The content is solely the responsibility of the authors and does not necessarily represent the official views of the aforementioned funding agencies.

\bibliography{refs/bibliography}

\end{document}